\documentclass[twocolumn,aps,prl,amsmath,showpacs]{revtex4}
\usepackage{graphicx}
\usepackage{dcolumn}%
\usepackage{bm}%
\usepackage{amssymb}
\usepackage{amsmath}
\usepackage{color}

\def\be{\begin{equation}}
\def\ee{\end{equation}}

\begin{document}

\title{Quantifying the Reversible Association of Thermosensitive Nanoparticles}

\author{Alessio Zaccone$^1$, Jerome J. Crassous$^2$, Benjamin B\'eri$^1$, and Matthias Ballauff$^3$}
\affiliation{${}^1$Department of Physics, Cavendish Laboratory, University of Cambridge, Cambridge CB3 0HE, United Kingdom}
\affiliation{$^2$Adolphe Merkle Institute, University of Fribourg, 1723 Marly, Switzerland}
\affiliation{$^3$Helmholtz Zentrum f\"ur Materialien und Energie, D-14109 Berlin, Germany, and Department of Physics, Humboldt-University, Berlin, Germany}
\date{\today}
\begin{abstract}
Under many conditions, biomolecules and nanoparticles associate by means of attractive bonds, due to hydrophobic attraction. Extracting the microscopic association/dissociation rates from experimental data is complicated by the dissociation events and by the sensitivity of the binding force to temperature ($T$). Here we introduce a theoretical model that combined with light scattering experiments allows us to quantify these rates and the reversible binding energy as a function of $T$. We apply this method to the reversible aggregation of thermoresponsive polystyrene/poly(N-isopropylacrylamide) core-shell nanoparticles, as a model system for biomolecules. We find that the binding energy changes sharply with $T$, and relate this remarkable switchable behavior to the hydrophobic-hydrophilic transition of the thermosensitive nanoparticles.


\end{abstract}
 \maketitle

\date{\today}

The association and self-organization of biomolecules~\cite{Stradner} and nanoparticles~\cite{Yan} play a fundamental role in many physiological processes~\cite{Morris}, such as protein amyloid formation which is responsible for neuro-degenerative diseases~\cite{Dobson}, as well as in technological processes. In particular the assembly of nanoparticles has become a key step in the synthesis of nanomaterials with new optical and mechanical properties~\cite{Daniel}. In order to understand and control all these association processes, it is essential to understand and control the association kinetics~\cite{Morris,Dobson}. This is challenging in the case of biomolecules and nanoparticles where often the microscopic binding energy, which directly determines the rate of dimer formation, is comparable to the thermal energy $k_{B}T$. This  poses two major difficulties: (i) the experimental detection of the binding energy has to be accurate down to the $k_{B}T$ scale, and (ii) the association kinetics is affected by dissociation events~\cite{Cellmer,Knowles} since the binding energy is comparable with the average kinetic energy of the particles/molecules. It has been recently shown that a clear understanding of protein aggregation under physiological conditions cannot be achieved without the understanding of aggregate dissociation~\cite{Knowles}.
A further complication arises from the fact that the binding energy can be very sensitive to changes of the solution parameters, such as $T$ and $pH$~\cite{Morris}.

In this Letter we propose a detection strategy that can overcome many of the difficulties to date in extracting quantitative information about the intrinsic rates of association and dissociation processes. The essence of the strategy is to combine a kinetic model with the analysis of dynamic light scattering (DLS) data of aggregation kinetics. Dynamic light scattering has proven a reliable tool for analyzing irreversible aggregation~\cite{holthoff}. Here we show that it can be used for measuring reversible processes as well, but the data analysis requires a new model fully accounting for dissociation processes. Here we introduce such a model and demonstrate its applicability in light scattering experiments of thermosensitive nanoparticles in aqueous suspension. We also show how to quantify the $T$-dependent interaction energy between the particles  once the rates are extracted.

\begin{figure}[h]
\includegraphics[width=0.8\linewidth]{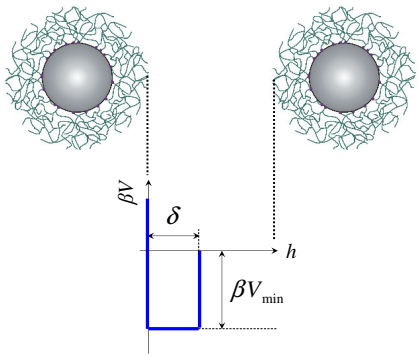}
\caption{(color online). Structure of the thermosensitive particles and of the effective square-well interaction, setting in at $T>T_{c}$, as a function of the surface separation $h$.} \label{fig1}
\end{figure}

Thermosensitive nanoparticles are interesting on their own and are the object of intense study because their size and interaction can be modified by changes of $T$ \cite{Yan}. They consist of a 52 $nm$ radius solid polystyrene core  onto which a polymeric network of crosslinked poly(N-isopropyl acrylamide) (PNIPAM) of $T$-dependent thickness ($\simeq50nm$ at $T<32^{o}C$ and $\simeq33nm$ at $T>32^{o}C$) is affixed (see Fig.\ref{fig1}). At room temperature this thermosensitive shell is swollen by the dispersing agent water, which is expelled if the suspension is heated above the critical temperature $T_{c}\!\!=\!\!32^{0}C$ of a volume transition. It is a well-established fact that these particles become attractive above $T_{c}$~\cite{Crassous06}. This is due to the hydrophobic interaction between the PNIPAM-network for which water has become a poor solvent under these conditions. As a consequence of this, aggregation sets in which is entirely reversible upon cooling.
The thermosensitive particles have been synthesized and characterized as described in \cite{Crassous06}. Core-shell particles with 5 $mol.$\% crosslinking are used in this study. In salt-free solution these particles have a good stability at all temperatures. Adding salt, however, screens the electrostatic repulsion and induces aggregation above $T_{c}$ \cite{Crassous06}. The reversible character of the aggregation is demonstrated in ~\cite{Suppl}. The kinetics was investigated by DLS using a ALV 5000 light scattering goniometer (Peters) at a scattering angle of 90$^0$ and wavelength $\lambda=632.8nm$. The experiment was done as follows: 2.3 $mL$ of the latex solution (either at $2.72\times10^{-3}$, $1.36\times10^{-3}$ or $0.27\times10^{-3}$ wt.\%) was equilibrated at the required temperature for 20 min. Then 0.2 $mL$ of a 0.625 $M$ solution of $KCl$ kept at the same $T$ were quickly added to induce aggregation. After homogenization the measurement was started and the average hydrodynamic radius $r_{h}$ of the colloidal constituents of the suspension was monitored as function of time (see Fig. 2). The time evolution of $r_{h}$ is linear in the early stage of association and for sufficient dilution and gives access to the association rate. If dissociation events occur on the time scale of observation, however, the measured association rate is an effective one (i.e. it contains the effect of dissociation and is therefore smaller than in the absence of dissociation). In the following we develop a model which allows us to account for this effect and to extract the microscopic dissociation rate from this effective association rate from the DLS experiments.\\

We start by considering the kinetics of reversible association between two particles $A$ to form a dimer $A_{2}$
\begin{equation}
A +A \rightleftharpoons A_{2}\label{eqmod1}
\end{equation}
The association rate be denoted by $k_{+}$ and the dissociation rate by $k_{-}$. If we denote with $n_{1}$ the concentration of monomers $A$ at time $t$ and with $N$ the total concentration of monomers at $t=0$, the evolution of $A$ is governed by the following equation
\begin{equation}
\frac{d n_{1}(t)}{dt}=-k_{+}n_{1}(t)^2+\frac{1}{2}k_{-}N-\frac{1}{2}k_{-}n_{1}(t)\label{eqmod2}
\end{equation}
where we made use of the conservation condition: $n_{2}(t)=(N-n_{1}(t))/2$, with $n_{2}$ the concentration of dimers $A_{2}$. With the initial condition $n_{1}(0)=N$, Eq.(2) admits the following solution:
\begin{equation}
n_{1}(t)=-\frac{k_{-}}{2k_{+}}+\frac{\mathcal{\sqrt{\mathcal{A}}}}{2k_{+}}\left[\frac{\tanh(\sqrt{\mathcal{A}}t/2)+\mathcal{B}/\sqrt{\mathcal{A}}}{1+(\mathcal{B}/\sqrt{\mathcal{A}})\tanh(\sqrt{\mathcal{A}}t/2)}\right]
\label{eqmod3}
\end{equation}
with $\mathcal{A}=k_{-}(k_{-}+4k_{+}N)$ and $\mathcal{B}=k_{-}+2k_{+}N$.
\begin{figure}[h]
\includegraphics[width=0.7\linewidth,angle=+90]{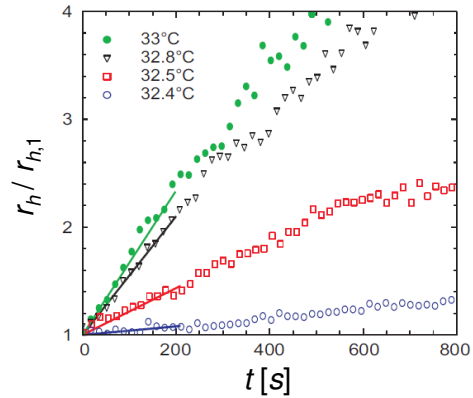}
\caption{(color online). Time-evolution of the average hydrodynamic radius of the colloidal suspension measured by DLS.} \label{fig2}
\end{figure}

According to~\cite{holthoff}, in the early stage of aggregation of Brownian particles where only monomers (1) and dimers (2) are present, the temporal evolution of the average hydrodynamic radius of the system as measured by dynamic light scattering (DLS) is given by
\begin{equation}
\frac{1}{r_{h}(t)}=\frac{I_{1}(q)n_{1}(t)/r_{h,1} + I_{2}(q)n_{2}(t)/r_{h,2}}{I_{1}(q)n_{1}(t)+I_{2}(q)n_{2}(t)}
\label{eqmod4}
\end{equation}
where $I_{1}(q)$ and $I_{2}(q)$ are the wave-vector ($q$)-dependent intensities of radiation scattered by monomers and dimers respectively, while $r_{h,1}=85nm$ and $r_{h,2}=1.38\times85nm$ are the hydrodynamic radii of monomer and dimer respectively~\cite{Suppl}. The temporal evolution of $r_{h}$ can be calculated by substituting Eq. (3) together with the conservation relation $n_{2}(t)=(N-n_{1}(t))/2$ into Eq.(4). The resulting expression can be expanded and as we are interested in the initial kinetic behavior, we can truncate to first order in $t$. This gives
\begin{equation}
{r_{h}(t)}\propto\frac{8N^{3}k_{+}^{3}I_{1}(q)I_{2}(q)r_{h,1}(1-r_{h,1}/r_{h,2})}{[2k_{-}+4k_{+}N-k_{-}(I_{2}(q)/I_{1}(q))(r_{h,1}/r_{h,2})]^{2}}t
\label{eqmod5}
\end{equation}
Upon taking the derivative and rearranging terms we obtain the standard form
\begin{equation}
\frac{1}{r_{h,1}}\frac{d r_{h}(t)}{dt}=\frac{I_{2}(q)}{2I_{1}(q)}\left(1-\frac{r_{h,1}}{r_{h,2}}\right)NK_{\mathrm{eff}}
\label{eqmod6}
\end{equation}
with the effective association constant taking account of reversibility given by
\begin{equation}
K_{\mathrm{eff}}=\frac{16k_{+}^{3}N^2}{[2k_{-}+4k_{+}N-k_{-}(I_{2}(q)/I_{1}(q))(r_{h,1}/r_{h,2})]^{2}}
\label{eqmod7}
\end{equation}
This equation is our key result which allows to extract the microscopic dissociation rate $k_{-}$ from the $K_{\mathrm{eff}}$  measured in the experiments.
One can verify by taking the $k_{-}=0$ limit of no dissociation that Eq.~\eqref{eqmod7} correctly recovers the well-known result for the \emph{irreversible} aggregation kinetics~\cite{holthoff}.
We access $K_\text{eff}$ experimentally by applying Eq.~(6) to data sets of the kind shown in Fig.2. In the analysis, $r_{h,1}$, $r_{h,2}$, $I_{1}(q)$, and $I_{2}(q)$ are all known parameters, as well as $k_{+}$ which for attractive colloids is given by the Smoluchowski diffusion-limited rate, $k_{+}= (8/3)k_{B}T/\eta$, with $\eta$ the water viscosity~\cite{Russel89}. (The details of the fitting procedure are given in~\cite{Suppl}.) The effective association constant is plotted in terms of the colloidal stability coefficient~\cite{Russel89}, $W_{\mathrm{eff}}=k_{+}/K_{\mathrm{eff}}$, as a function of $T$ in the left inset of Fig. 3. It is seen that it decreases very sharply around $T_{c}$, after which it reaches a diffusion limited-like plateau.


The knowledge of $k_{-}$ also allows us to estimate the energy scales for attraction between the nanoparticles, as we now explain. The rate at which a dimer dissociates into two monomers must coincide with the rate at which one particle escapes from the attractive potential well that binds it to a second particle. The escape is a stochastic process promoted by the Brownian motion in competition with the attractive interaction~\cite{kramers}. In order to keep the treatment analytical we assume an effective attractive square-well potential of width $\delta$ and depth $V_{\mathrm{min}}$. Such an effective potential is schematically depicted in Fig.\ref{fig1}. The approximate Kramers formula for the rate of escape from a square-well potential gives~\cite{egelhaaf}
\begin{equation}
k_{-}=\frac{D}{\delta^{2}}e^{-V_{\mathrm{min}}/k_BT},
\label{eqmod8}
\end{equation}
where $D=k_{B}T/6\pi\eta r_{h,1}$ is the diffusion constant ($D\simeq3.4\times10^{-12} m^{2}/s$ at $T\simeq32^{o}C$, in our system). 
To estimate $\delta$, note that the attraction between the particles is due to  the hydrophobic interaction between the PNIPAM microgel grafted layers. As shown in many studies in the past, the typical range of the attractive interaction between two organic hydrophobic surfaces in water is about $10nm$~\cite{israelachvili}. As this value is mostly independent of the surface chemistry~\cite{israelachvili}, we take it here as the width $\delta$ of the effective well. Furthermore, this value is much larger than the molecular roughness of the PNIPAM shell, which is of the order of $1nm$~\cite{Suppl,Crassous06}, such that the attractive hydrophobic force dominates over repulsive entropic protrusion forces which are much shorter-ranged~\cite{israelachvili}. There is now only one free parameter in our model, the depth of the effective attraction well, $V_{\mathrm{min}}$, i.e., the binding energy of the dimers. One-parameter fits to the measured values of $W_\text{eff}$ reveal that  $V_{\mathrm{min}}$ has an almost step-like behavior, as shown in the right inset of Fig. 3. It increases sharply from zero in a narrow $T$ interval and saturates at $V_{\mathrm{min}}^{\infty}=12k_{B}T$.

\begin{figure}[h]
\includegraphics[width=0.8\linewidth]{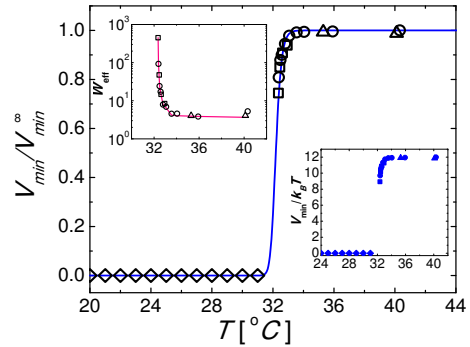}
\caption{(color online). Main frame: The experimental $T$-dependence of the binding energy (symbols) together with the theoretical model given by Eq. (13) (line). Triangles: $2.5\times10^{-4}$. Squares: $1.25\times10^{-3}$ wt\%. Circles: $2.5\times10^{-3}$ wt\%. Insets: left-the one-parameter fit to experimental measurements of $W_{\mathrm{eff}}$ as a function of temperature. Triangles: $2.5\times10^{-4}$. Squares: $1.25\times10^{-3}$ wt\%. Circles: $2.5\times10^{-3}$ wt\%. Line: theoretical fit varying $V_{\mathrm{min}}$ as the unique fitting parameter. The experiments were done using the methods of Ref.\cite{holthoff}; right-the average binding energy as a function of $T$.} \label{fig3}
\end{figure}

The observed sharp behavior of $V_\text{min}$ can be related to the $T$-dependent behavior of thermoresponsive nanoparticles. In the $T$ interval where the turnover in $V_{\mathrm{min}}$ occurs, the PNIPAM network grafted on the particle surface undergoes a transition from a hydrophilic (L) to a hydrophobic (B) state with increasing $T$. The transition is associated with a volume change from $V_L$ to $V_B$. In the thermodynamic limit, this would be a first-order phase transition~\cite{Kubo} with the volume as the order parameter, but now it is smeared due to the finite size of a nanoparticle~\cite{Note}. On a phenomenological level, this transition is captured by a partition function
\begin{equation}
Y(T,P)=\int dV \exp[-g(T,P,V)/k_BT],
\end{equation}
where $g(T,P,V)$ plays the role of a Landau free energy, accounting for the many states of the nanoparticle realizing volume $V$. (We use the $T-P$ ensemble since the externally imposed parameters are the temperature and the pressure $P$~\cite{Kubo}.) The first order-like transition means that $g(T,P,V)$ has sharp minima at $V_L$ and $V_B$, which are degenerate at the transition temperature $T_c$. The partition function can be thus approximated as
\begin{equation}
Y(T,P)=Y_L(T,P)+Y_B(T,P),
\end{equation}
where $Y_\alpha(T,P)=\int_{V\approx V_\alpha} \exp[-g(T,P,V)/k_BT]$. The probability that the particle is in the $\alpha$ state is \mbox{$p_\alpha\!=\!Y_\alpha/Y$}. In a narrow temperature range $|T-T_c|\ll T_c$, the restricted partition functions $Y_\alpha$ can be written as
\begin{equation}
Y_\alpha(T,P)=\exp\left(-\frac{1}{k_BT}[H_\alpha(T_c)-TS_\alpha(T_c)]\right).
\label{eq:Yres}
\end{equation}
Since the width of the transition is only few Kelvins, Eq.~\eqref{eq:Yres} applies in the whole temperature range from $p_L=1$ to $p_B=1$. This allows us to identify $H_\alpha(T_c)$ and $S_\alpha(T_c)$ as the enthalpy and entropy of the nanoparticle in the $\alpha$ state away from the transition region. In terms of these quantities, the probability of the hydrophobic state is
$p_B=(1+\exp[(\Delta H-T\Delta S)/k_BT])^{-1}$,
where $\Delta H=H_B-H_L>0$ and $\Delta S=S_B-S_L<0$ characterize the change in the enthalpy and the entropy across the transition. At $T_c$, one has $Y_L=Y_B$ which implies $\Delta H=T\Delta S$. Note that the probability distribution
\begin{equation}
p_B=\frac{1}{1+\exp[-(T_c-T)\Delta S/k_BT]},
\label{eq:pB}
\end{equation}
has the same form as for a two level system, $p_B=[1+\exp(\varepsilon/k_BT)]^{-1}$ with a $T$-dependent level splitting $\varepsilon=-(T_c-T)\Delta S$.

In a dilute solution, $p_B$ gives the proportion of hydrophobic nanoparticles. To relate Eq.~\eqref{eq:pB} to the experiment, we have to now combine the above statistical mechanics model with some assumptions about the initial aggregation. We base these on the following properties: in the hydrophilic state, the particles repel each other near contact due to combination of steric and hydration forces, while two particles that are both in the hydrophobic state attract each other via the hydrophobic force.
In analogy with the kinetic model, we assume that collisions between two hydrophobic particles happen with the binding energy $V_\text{min}^\infty$, and take all other collisions to be of hard sphere type. Starting from all nanoparticles in a monomer state, if after a short time there were $\alpha$ binary collisions, $p_B^2\alpha$ of them were binding. (We always focus on the early stage of aggregation in very dilute systems where binary collisions are dominant.)
In the kinetic model, the number of short time collisions is just the number of dimers $n_2$ at short times; in analogy we take $\alpha=n_2$. Now requiring that the short time total binding energies $n_2 p_B^2(T) V_\text{min}^\infty$ and $n_2 V_\text{min}(T)$ of the statistical mechanics and kinetic models are the same gives
\begin{equation}
\frac{V_\text{min}(T)}{V_\text{min}^\infty}=\left(\frac{1}{1+\exp[-(T_c-T)\Delta S/k_BT]}\right)^{2}.
\label{eq:VS}
\end{equation}
Eq.~\eqref{eq:VS} is compared with the experimental data in Fig. (3). The comparison confirms that $T_{c}=32+273.16 K$ and allows for estimating $\Delta S\approx-1420~k_{B}$.
This can be interpreted as the interfacial hydration entropy loss when the microgel layer turns from hydrophilic into hydrophobic. The corresponding energy is
$-T_c\Delta S=3.62\times 10^{6}\mathrm{J mol^{-1}}$. The hydration entropy loss per monomer was measured for free PNIPAM chains in water and is $\sim10 \mathrm{J K^{-1}mol^{-1}}$~\cite{kunugi}. Thus, if chains were to behave as free chains, we would have $\sim10^{3}$ equivalent monomers per particle. However, chains in the microgel layer are grafted and crosslinked. The typical entropy change for deformed grafted chains is $\sim10^{3}$ times smaller than for free chains~\cite{Ohno}. The estimate order of $\sim10^{6}$ monomers in the grafted layer agrees with the value $\sim3\cdot10^6$ based on the known composition of our system~\cite{Crassous06}.

To conclude, we have presented a kinetic model that can account for spontaneous dimer dissociation which, combined with light-scattering experiments, provides a robust protocol for quantifying the dissociation rate and the reversible association (dimerization) energy of 
mutually attractive Brownian molecules and nanoparticles.
We have demonstrated our method on PNIPAM-coated core-shell thermosensitive nanoparticles where the attraction stems from the hydrophobic effect. We found that the association energy  changes sharply from zero to \mbox{$\sim12k_{B}T$} across the transition temperature for the swelling of the PNIPAM shell. Using a statistical mechanical analysis, we have related this behavior to the two-level nature of the nanoparticles.
Our method opens up the perspective of systematically quantifying the effective hydrophobic attraction~\cite{Chandler}, between complex biomolecules, such as proteins. This has potential applications to other problems where dissociation events are important, such as the amyloid formation~\cite{Knowles}, responsible for neuro-degenerative diseases~\cite{Dobson}.

We acknowledge financial support by the Swiss National Science Foundation, grant $PBEZP2-131153$ (A.Z.), by the Adolphe Merkle Foundation (J.C.),by EPSRC Grant EP/F032773/1 (B.B.), and by the Deutsche Forschungsgemeinschaft (M.B.).


\begin{thebibliography}{99}
\bibitem{Note} The transition from $V_{L}$ to $V_{B}$ is superimposed on a much slower linear volume decrease which is unrelated to the hydrophobicity, hence it can be neglected in our analysis.
\end{thebibliography}
\end{document}